\documentclass[runningheads]{llncs}
\usepackage{graphicx}
\usepackage{url}
\begin{document}
\title{The World Literature Knowledge Graph}

%
%\titlerunning{Abbreviated paper title}
% If the paper title is too long for the running head, you can set
% an abbreviated paper title here
%
\author{Marco Antonio Stranisci\inst{1}\orcidID{0000-0001-9337-7250}\thanks{Contributed equally to this work} \and
Eleonora Bernasconi\inst{2} \orcidID{0000-0003-3142-3084}\inst{*} \and
Viviana Patti\inst{1}\orcidID{0000-0001-5991-370X} \and
Stefano Ferilli\inst{2}\orcidID{0000-0003-1118-0601} \and
Miguel Ceriani\inst{2} \orcidID{0000-0002-5074-2112} \and
Rossana Damiano\inst{1}\orcidID{0000-0001-9866-2843}}

%Eleonora Bernasconi \orcidID{0000-0003-3142-3084}
%Miguel Ceriani \orcidID{0000-0002-5074-2112}
%
\authorrunning{Stranisci et al.}
% First names are abbreviated in the running head.
% If there are more than two authors, 'et al.' is used.
%

%\institute{Università degli Studi di Bari Aldo Moro, Bari, Italy \email{\{name,surname\}@uniba.it}}

\institute{University of Turin. Corso Svizzera 185, Turin, Italy \and
University of Bari Aldo Moro. Via Orabona 4, Bari, Italy
\email{\{marcoantonio.stranisci,viviana.patti,rossana.damiano\}@unito.it}\\
\email{\{eleonora.bernasconi,stefano.ferilli,miguel.ceriani\}@uniba.it}}
\maketitle              % typeset the header of the contribution
\begin{abstract}
Digital media have enabled the access to unprecedented literary knowledge. Authors, readers, and scholars are now able to discover and share an increasing amount of information about books and their authors. However, these sources of knowledge are fragmented and do not adequately represent non-Western writers and their works. In this paper we present The World Literature Knowledge Graph, a semantic resource containing $194,346$ writers and $965,210$ works, specifically designed for exploring  facts about literary works and authors from different parts of the world. The knowledge graph integrates information about the reception of literary works gathered from $3$ different communities of readers, aligned according to a single semantic model.
The resource is accessible through an online visualization platform, which can be found at the following URL: \url{https://literaturegraph.di.unito.it/}\footnote{Credentials: email: iswc2023@wlkg.com - password: iswc2023}. This platform has been rigorously tested and validated by $3$ distinct categories of experts who have found it to be highly beneficial for their respective work domains. These categories include teachers, researchers in the humanities, and professionals in the publishing industry. The feedback received from these experts confirms that they can effectively utilize the platform to enhance their work processes and achieve valuable outcomes.
%The resource is accessible through an online visualization platform (\url{https://literaturegraph.di.unito.it/}) that was tested by $3$ categories of experts that can profitably exploit it in their work: teachers, researchers in the humanities, and people working in the publishing industry. 

\end{abstract}

\section{Introduction}
The impact of digital media on the literary ecosystem has led to a transformation of reading \cite{nakamura2013words} and researching \cite{o2015only} practices.
Digital media represent an unprecedent opportunity for studying the World Literature \cite{damrosch2003world}. Digital platforms are not only open windows on different parts of the world, but also privileged viewpoints on how communities of readers receive and share literary works \cite{stranisci2023user}. Such a transformation also affected research practices, which benefit from the web as a source of knowledge.

These opportunities are however limited by a series of issues. The knowledge stored in these archives is vast, but fragmented: only a minimal part of writers and works is mapped from one source to another and many of them do not rely on a semantic model. This hinders the study of the writers and their reception by different groups of readers. Furthermore, it has been proved that some of these resources are characterized by the underrepresentation of non-Western people. It is the case of Wikidata \cite{adams2019counts} and Wikipedia \cite{field2022controlled} that are both affected by gender and an ethnic bias. 

The World Literature Knowledge Graph (WL-KG) is a knowledge base developed for tackling these issues. The resource includes $194,346$ writers and their works gathered from three sources of knowledge: Wikidata, Open Library, and Goodreads. Such a collection relies on a common ontology network \cite{stranisci2021representing} specifically developed with the aim of emphasizing the ethnic origin of writers and the readers' response about them and their works.

The WL-KG is intended to support two main types of tasks: (i) the analysis of the underrepresentation of non-Western writers; (ii) the reception of works by different communities of readers. These  tasks, in turn, can support the implementation of  applications like recommender systems \cite{rajpurkar2015book}, and discovery tools \cite{polley2020simfic}, which  may take advantage from the more balanced representation of literary world provided by the knowledge base. The WL-KG is also intended as a tool for all professionals that work in the literary field (e.g.,  researchers in the humanities and publishers) and operate in multicultural contexts (e.g., teachers, educators, activists). In order to make this resource accessible to these target, it is hosted on a visualization platform \cite{Bernasconi2023} that allows for a graph-based exploration of the KG. Both the platform and the WL-KG were tested by $3$ categories of experts who evaluated them along three dimensions: completeness, accuracy, and usability. Results showed that our resource may be considered as an alternative to traditional literary search tools, especially for the discovery of new writers.

This paper is structured as follows. In Section \ref{sec:related} related work and theoretical background are presented. Section \ref{sec:semantic_model} describes the semantic model on which the WL-KG relies. Section \ref{sec:kg}  describes the creation of the resource, while Section \ref{sec:visualization} illustrates its  implementation in a visualization platform.  Section \ref{sec:evaluation} reports on the evaluation of the resource.

%Fotografia della letteratura che non è tarata su un modello occidentale, ma che si basa su standard internazionali e focalizzato sulle comunità di lettori. (Reception theory)
\section{Background and Related Work} \label{sec:related}
In this section we first briefly describe the World Literature theoretical framework. Then, we  review the related work in two fields: semantic resources designed for literary studies and linked data visualization platforms.

\subsection{Theoretical Framework}
World Literature is a recent approach to literary studies that emphasizes the idea of works as windows on different parts of the world \cite{damrosch2003world}. In such a perspective, national and chronological boundaries must be overcome and a crucial step of the analysis is how works transcend their local contexts to be globally received \cite{jauss1970literary,benwell2012postcolonial}. Such a framework gained prominence in last years with the availability of an unprecedented knowledge about writers and their works enabled by social media: this paved the way for the development of distant reading approaches \cite{moretti2000conjectures} as well as digital humanities studies of digital platforms \cite{hube2017world}. 
The centrality of reception and the emphasis on a non-Western centric approach are two features from this theory that was adopted for modeling the WL-KG. In fact, our resource can be used not only for discovering writers and works from the world, but also to analyze how communities of readers increase or decrease their underrepresentation, and to devise way to contrast it.

\subsection{Semantic Technologies for Literary Studies}
Several digital resources that provide information about literary works and writers are available online. Wikidata \cite{vrandevcic2014wikidata} 
%which is a KG built upon Wikipedia (non è vero, quello è dbpedia)
is a general-purpose KG which includes knowledge about writers and their works. Other archives are domain-specific: Goodreads is a site owned by Amazon where readers share their impressions about books. Open Library is a project of the Internet  Archive\footnote{\url{https://archive.org/}}  where users can borrow books. Among these three archives, only Wikidata relies on the Linked Open Data paradigm. Open Library exposes its data through APIs, while Goodreads dismissed its APIs in $2020$. This leads to issues in data gathering and mapping, since there is no unified model to align these resources.

Some digital archives are monographic and curated by teams of experts. It is the case of The European Literary Text Collection\footnote{https://www.distant-reading.net/eltec/} \cite{schoch2018distant}, a multi-lingual dataset of novels written from 1848 to 1920; DraCor\footnote{https://dracor.org/}  \cite{fischer2019programmable}, a collection of plays corpora in multiple languages; MiMoText\footnote{https://mimotext.github.io/}, a parallel corpus of French and German novels published from $1750$ to $1799$. 

Other resources are more oriented to explore the intersection between people and society. The Japanese Visual Media Graph\footnote{https://jvmg.iuk.hdm-stuttgart.de/} \cite{pfeffer2019japanese} gathers data about Japanese visual media (including manga and visual novels) from communities of fans. The Orlando Textbase\footnote{https://www.artsrn.ualberta.ca/orlando/} \cite{simpson2013xml} is a KG developed for exploring feminist literature. WeChangeEd\footnote{https://www.wechanged.ugent.be/} \cite{van2021women} is a KG of $1,800$ female editors born between 1710 and 1920, aligned with Wikidata.

The WL-KG is the first resource designed to study the intersection between literary production and ethnic information about writers. There are research projects that analyze the world of literature according to Wikipedia \cite{hube2017world}, but this is the first attempt to release a resource which could be at the same time a platform to foster
%DH
digital humanities
and literary studies and a benchmark dataset for analyzing the knowledge gaps that affect an authoritative source like Wikidata in the literary domain.

\subsection{Visualization Platforms}
Many works deal with interfaces for visualising linked data \cite{desimoni2020,po2020,queryvowl_2,webvowl,rdf4u,rdfdigest+,relfinder,tarsier}, but only some focus on exploring and disseminating domains related to digital humanities, primarily digital libraries \cite{bernasconi2023linked,fenlon2017researchspace}. The interaction paradigm and the information reduction strategies are the two main characteristics of an interface for visualising linked data.

%For example,
ARCA~\cite{Bernasconi2023} is a modular system that deals with knowledge extraction from a digital library,
%collaborative
visualisation,
and collaborative validation of  automatically extracted associations between concepts and books~\cite{bernasconi2022automatic}.
ARCA uses two different interaction paradigms:
%the first is
the node-link paradigm for visualising resources extracted and linked to the DBpedia knowledge base \footnote{\url{https://dbpedia.org/}}, and
%. At the same time, the second is
the tabular paradigm for the visualisation of additional metadata related to books.
As an information reduction strategy, ARCA allows for incremental visualisation of resources.

On the other hand, Yewno Discover \cite{bolina2019yewno} allows node-link visualisation of concepts contained in a digital library. Unlike ARCA, Yewno has a static and non-incremental visualisation of resources but uses ranking algorithms to filter the displayed content.

Another tool is ResearchSpace \cite{fenlon2017researchspace}, an open-source platform that facilitates working with digital cultural heritage data in a linked data environment, enabling improved discoverability and reuse of data. The platform includes a node-link interaction paradigm, which employs incremental visualization for knowledge exploration. Additionally, it allows for collaborative annotation of texts or images.

Thanks to the flexibility and modularity of the ARCA system, we have chosen to build upon it by creating an extension called SKATEBOARD (Semantic Knowledge Advanced Tool for Extraction Browsing Organization Annotation Retrieval and Discovery), as described in Section \ref{visualization_platform}. This extension has been customized and updated to meet the specific needs
of users interacting with
% of our
the World Literature Knowledge Graph. % work.

\section{The Semantic Model} \label{sec:semantic_model}
In this section the semantic model adopted for the WL-KG is described. After a general introduction of the model and the authoritative ontologies to which it is aligned, we focus on two aspects that we modeled through our ontology network: the interaction between writers and their ethnic origin and the representation of the publishing history of the works.

\subsection{The UR-Ontology Network}
\begin{sloppypar}
The ontology network serves two main functions: modeling ethnic-based underrepresentation of writers; mapping different digital libraries under a unique data model.
Data in the WL-KG are modeled according to the Under-Represented Ontology Network (UR-O) composed of two modules: a revised version of the Under-Represented Writers Ontology (URW-O)\footnote{\url{https://purl.archive.org/urwriters/lode}} \cite{stranisci2021representing} and a module for the encoding of works: the Ontology of Under-Represented Books
(URB-O)\footnote{\url{https://purl.archive.org/urbooks/lode}}. 

The ontology network is mapped onto three authoritative ontologies: the Functional Requirements for Bibliographic Records (FRBR)~\cite{tillett2005frbr}, the PROV Ontology (PROV-O)~\cite{lebo2013prov}, and the Descriptive Ontology for Linguistic and Cognitive Engineering (DOLCE)~\cite{gangemi2002sweetening}. FRBR is a standard for modeling the relationship between a work (\textsc{frbr:Work}), its expressions (\textsc{frbr:Expression}), and manifestations (\textsc{frbr:Manifestation}). From PROV-O the relationships of attribution, association, and derivation are inherited, in order to make explicit the sources from which data were gathered (\textbf{prov:wasDerivedFrom}), the people and organizations involved in specific editions of given works (\textbf{prov:wasAssociatedWith}) and their roles (e.g., publisher, translator), and the attribution of a work to its creator (\textbf{prov:wasAttributedTo}). DOLCE has been used as a reference model for encoding biographical and publishing events, which are represented as time-bounded perdurants in which entities play specific roles. This allows representing publications as an event where a set of entities participates (\textbf{dul:hasParticipant}) and life events (e.g., Birth, Migration) as situations which are setting for (\textbf{dul:isSettingFor}) agents and their roles. 
%\\Our ontology network serves two functions: modeling ethnic-based underrepresentation of writers; mapping different digital libraries under an unique data model. 
\end{sloppypar}
\begin{figure}[t]
    \centering
    \caption{An example of how the concept of `Transnational' writer is encoded in our semantic model}
    \includegraphics[width=\textwidth]{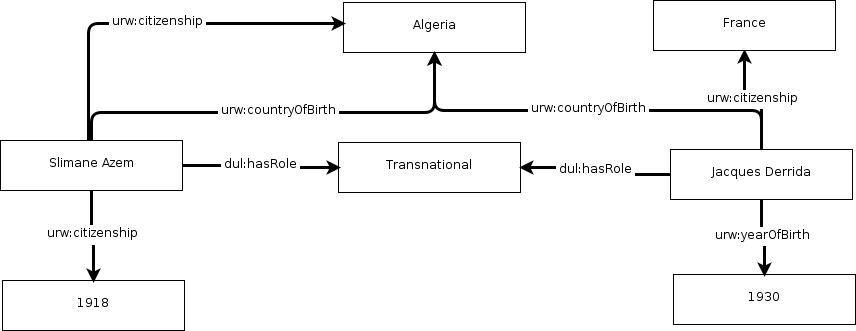}
    
    \label{fig:example_transnational}
    \vspace{-1mm}
    
\end{figure}
\subsection{Modeling Underrepresentation} For modeling ethnic-based underrepresentation of writers we relied on two criteria derived from post-colonial studies \cite{spivak2003can}. To be potentially under-represented an author must either (i) be born in a non-Western former colony country or (ii) belong to an ethnic minoritiy in a Western country. Using the country of birth as a criterion is prone to false positives though, since many writers with Western origin were born in former colonies (e.g., George Orwell, Rudyard Kipling). In order to mitigate such issue we chose to adopt the term `Transnational', which is broader than `Under-Represented’ since it refers to people who ``operated outside their own nation’s boundaries, or negotiated with them'' \cite{boter2020unhinging}. Furthermore, we classified as `Transnational’ only people born in former colonies from Latin America and Caribbeans since 1808, and in former African and Asian colonies since 1917,
%for reducing
to reduce
the number of people of Western origin selected by
%associated with
this condition. The first date marks the beginning of the Spanish American wars of independence; the second was chosen as a symbolic beginning of the decolonization process in Africa and Asia. Finally, we coupled the condition of being `Transnational’ with the citizenship of an author in order to reveal potentially false under-represented writers which may be still present in the knowledge base. As it can be observed in Figure \ref{fig:example_transnational}, Jacques Derrida and Slimane Azem are both classified (\textbf{dul:hasRole}) as `Transnational' in the KG, since they were born in Algeria, a former African colony, in $1918$ and $1930$. The specification of their citizenship (\textbf{urw:citizenship}) provides additional information about Jacques Derrida, who was not an Algerian citizen despite Algeria is its country of birth. This allows users to infer his European origins.

\subsection{Modeling Works Publishing History}
\begin{figure}[t]
    \centering
    
    \includegraphics[width=1.0\textwidth]{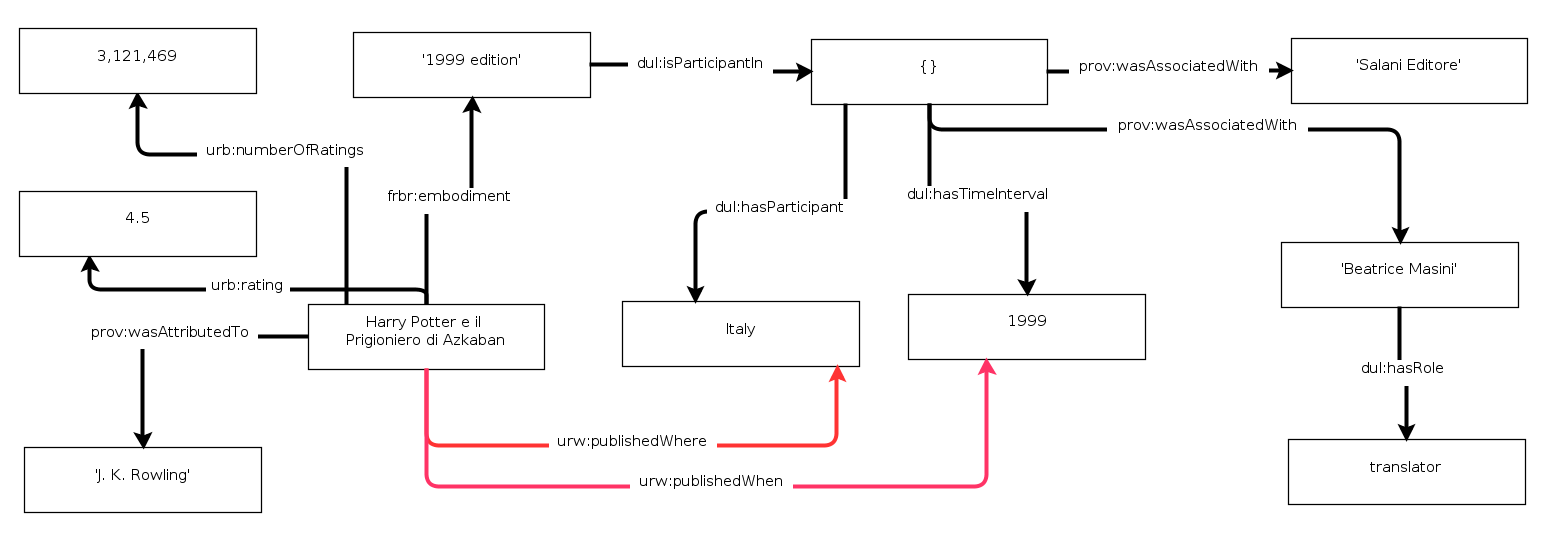}
    \caption{An example of publication}
    \label{fig:publication}
    \vspace{-1mm}
    
\end{figure}

Before gathering data from Wikidata, Open Library, and Goodreads we designed a common data model for aligning literary information that %is differently shaped by
the platform
represents in heterogeneous shapes. Following the FRBR ontology, we defined each work in the platform as an instance of type \textsc{frbr:Expression}, which is described as the ``intellectual or artistic realization of a work in the form of alpha-numeric, musical, or choreographic notation''.
We then defined the concept of \textsc{Edition} as a subclass of \textsc{frbr:Manifestation}, namely ``the physical embodiment of an expression of a work''. These two concepts are linked through the property \textbf{frbr:embodiment}. Such semantic relationship is wrapped in a \textsc{urb:Publication} pattern, which is a subclass of a \textsc{dul:Event}.
An event in DOLCE can be used as a reification
% for providing
to provide
richer descriptions of a property. In our case this type of pattern is adopted for two reasons: (i) expressing a large number of facts about an edition (place, date, language of publishing and publisher) in a compact way; (ii) encoding roles of people who contributed to a publication without being the author of a work.
A final feature of the semantic model is the reception of works from communities of readers. Depending on the source of knowledge from which a work is derived, it may have an average rating (\textbf{urb:rated}), a number of ratings (\textbf{urb:numberOfRatings}), or a number of readers (\textbf{urb:numberOfReaders}). Figure \ref{fig:publication} shows an example of our representation of works. `Harry Potter e il Prigioniero di Azkaban', namely the Italian version (\textsc{frbr:Expression}) of the 3rd Harry Potter book, \textbf{prov:wasAttributedTo} to J. K. Rowling, has an average rating and a number of ratings from the Goodreads community, and it has as \textbf{frbr:embodiment} the `$1999$ edition'. The latter in turn participates (\textbf{dul:isParticipantIn}) to a \textsc{urb:Publication}, a blank node entity that can be used for expressing several information: country of publication, year of publication, publisher, and translator. The translator is linked to the publication through the property \textbf{prov:wasAssociatedWith} and \textbf{dul:hasRole} `translator'.
Such %A similar
representation supports a thorough exploration of the intersections between writers' biographies and their publishing history as well as a more accurate analysis of their relationships with other authors and people working in the publishing industry.
It is however a
%redundant
verbose
encoding that may affect the usage of this resource.
In order to avoid this issue, we defined a set of property chains that directly link works to bibliographical information. Examples of these properties are shown in red in Figure~\ref{fig:publication}.

\section{Creation of the WL-KG} \label{sec:kg}
In this section, we describe the process involved in creating our World Literature Knowledge Graph (WL-KG), which can be queried online through the sparql endpoint available at \url{https://kgccc.di.unito.it/sparql/wl-kg}. We first describe our strategy for mapping knowledge from Wikidata onto Open Library and Goodreads. We then introduce our strategy for evaluating the mapping. Finally, we provide some statistics about the number of literary facts collected from each platform and about the interaction of communities of readers with works.

\subsection{Mapping between Platforms}
The data collection process started from Wikidata. From this knowledge base we gathered all the $194,346$ entities of type Person (wd:Q5) with occupation (wdt:P106)  writer (wd:Q36180), novelist (wd:Q6625963), or poet (wd:Q49757) born after $1808$ and having information about their place of birth. For each author, we collected the ethnic group, gender, date and place of death, Wikipedia page, and all the works associated with them. 

%Since the aim of the resource is to provide a snapshot of the World Literature as it is depicted by different communities of readers, the data collection process focused on user-generated content platforms rather than institutional initiatives. However, a quantitative analysis of external identifiers in Wikidata showed a low percentage of authors linked to these platforms and a high percentage of links to authoritative sources like the Virtual International Authority File Name (VIAF)

%For enriching
To enrich
the knowledge base we first conducted a quantitative analysis of their external identifiers in writers Wikidata pages. We focused on three of them: writers' Virtual International Authority File Name (VIAF) ID, Open Library ID and Goodreads ID. A fourth platform, Library Things, was not included in the data collection process given the low number of links from Wikidata and the impossibility of automatically obtaining authors IDs from that website. In Table~\ref{table:identifiers} it is possible to observe that the $84\%$ of writers has a VIAF ID, the $18.5\%$ an Open Library ID, and the $4.5\%$ a Goodreads ID. In order to increase the percentage of writers mapped to VIAF and Open Library identifiers, we adopted three heuristics: %for retrieving such information:
\begin{itemize}
    \item We retrieved all the names of the writers through the OpenLibrary APIs and kept only the entities fulfilling two conditions: (a) an exact string match between the author name in our KG and the one in OpenLibrary; (b) the same year of birth in our KG and in OpenLibrary. As a result, we obtained $19,737$ additional ids.
    \item We scraped all writers' names from Goodreads sitemap\footnote{\url{https://www.goodreads.com/siteindex.author.xml}} filtering out all homonyms. We then mapped all the names in our KG onto Goodreads author list, keeping only the string matches. We thus obtained $26,019$ new ids.
    \item We searched all ISBNs related to each authors through VIAF and performed a search through ISBN on Open Library and Goodreads, that allowed retrieving $22,661$ Open Library IDs and $44,142$ Goodreads IDs.
    
\end{itemize}

\subsection{Quality Assessment of the Mapping}
\begin{figure}[ht]
    \centering
    \includegraphics[width=\textwidth]{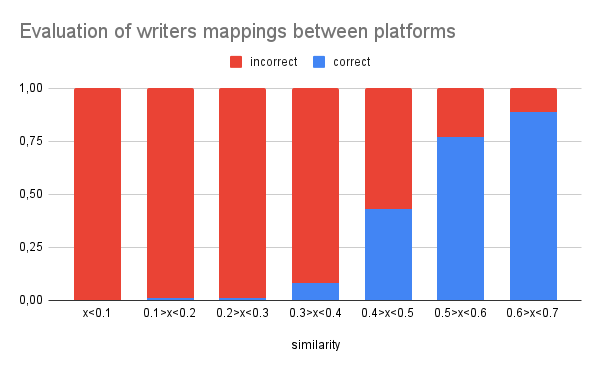}
    \caption{Results of the evaluation of writers mappings between Wikidata, Goodreads, and Open Library}
    \label{fig:mapping_eval}
    \vspace{-1mm}
    
\end{figure}

After the mapping, we performed a quality assessment of a sample of links between Wikidata and Goodreads, and between Wikidata and Open Library for removing incorrect links before gathering works. Our evaluation strategy is composed of three steps.
We computed the gestalt pattern similarity \cite{ratcliff1988pattern} between the names of the same writer in different platforms.
For instance, Esther Salamanwiki\footnote{\url{https://www.wikidata.org/wiki/Q4405658}}
is linked to her Goodreads page\footnote{\url{https://www.goodreads.com/author/list/618352.Esther_Polianowsky_Salaman}},
where she is referred as `Esther Polianowsky Salaman'.
The two strings have a gestalt pattern score \cite{gestaltpattern} of $0.7$. Then, we manually checked random samples of $100$ name pairs with $7$ degrees of similarity: $x<0.1$, $0.1\ge x<0.2$, $0.2\ge x<0.3$, $0.3\ge x<0.4$, $0.4\ge x<0.5$, $0.5\ge x<0.6$, $0.6\ge x<0.7$. As it can be observed in Figure \ref{fig:mapping_eval}, the percentage of correct links is directly proportional to the similarity between the name by which the writer is referred to in different platforms. In particular, the accuracy dramatically increases with a similarity between $0.5$ and $0.6$ ($77\%$ of correct links) reaching a $89\%$ of accuracy with a similarity between $0.6$ and $0.7$. 

Finally, we set a similarity threshold for filtering out potentially incorrect links. In order to privilege precision over recall, we set the threshold at $0.7$.  
As a final result, we obtained $71,706$ ($36.8\%$) writers with an Open Library ID and $79,158$ ($40.7\%$) with a Goodreads ID (Table \ref{table:identifiers}). The percentage of writers linked to at least one of the two platforms is $54\%$.

\begin{table}[ht]
\centering

\begin{tabular}{l|r|r}
    \textbf{Identifier} & \textbf{Before Mapping} & \textbf{After Mapping} \\
    \hline
    VIAF & $163,353$ ($84\%$) \\
    Open Library & $36,097$ ($18.5\%$) & $71,706$ ($36.8\%$)\\
    Goodreads & $8,997$ ($4.6\%$) & $79,158$ ($40.7\%$)\\
    
\end{tabular}
 \caption{Number of authors with an external identifier}
 \label{table:identifiers}
\end{table}

\subsection{Data Collection and Statistics}

After the augmentation of external identifiers of authors, we collected all their works in these platforms. OpenLibrary APIs allows retrieving all works, and for each work it is possible to obtain all editions. Results include a set of useful publishing information, readers count, ratings, and number of ratings. Goodreads does not provide APIs, but allows for web scraping. Hence, we first collected the list of all works from writers pages, their ratings and number of ratings, then we obtained publishing information through Google Books APIs.

In order to emphasize the role of readers communities, we only kept works that had received at least one reception or that were marked as read by at least one user. Table \ref{table:works} shows the number of works collected from each platform and the number of writers associated with at least one work from them. As it can be observed, Goodreads includes a higher number of works and writers with at least one work. Furthermore, both Open Library and Goodreads show a higher percentage of `Transnational' writers than Wikidata: $12.6\%$ and $11\%$ against $8.6\%$.

\begin{table}[ht]
\centering

\begin{tabular}{l|p{4cm}|p{3cm}}
    \textbf{Source} & \textbf{Writers with at least 1 work (\% Transnational)} & \textbf{N. of works} \\
    \hline
    Wikidata & $22,515$ ($8.6\%$) & $117,798$ \\
    Open Library & $24,370$ ($12.4\%$) & $226,108$\\
    Goodreads & $60,201$ ($11.0\%$) & $627,214$ \\
    Total & $71,443$ ($10.6\%$) & $971,120$ \\
    
\end{tabular}
 \caption{Number of works for each platform}

 \label{table:works}
\end{table}

%Wikidata is a knowledge base built over an ontology. Open Library allows accessing data in json format, distinguishing between works and editions, but it does not rely on a semantic model. Goodreads dismissed its APIs in 2020, thus not providing structured knowledge.

The analysis of readers communities may also be observed through the lens of the number of interactions between readers and works. While Wikidata does not include users evaluation of literary works, it is possible to obtain this information from Goodreads and Open Library. Both expose the number of ratings and the average rating, while the latter also exposes the number of readers. Table \ref{table:readers} shows the number of interactions between readers and literay works in the two platforms. As it can be observed, absolute numbers are incomparable: there are $112.708$ ratings in Open Library against $1.7$ billions in Goodreads. The percentage of ratings about Transnational works is higher on Open Library ($6\%$) than in Goodreads ($4.9\%$), while both platforms show a slightly higher average rating of Transnational writers.

\begin{table}[ht]
\centering

\begin{tabular}{l|l|l|l}
    \textbf{Source} & \textbf{Average rating} & \textbf{N. of works} & \textbf{N. of readers}\\
    \hline
    
    Open Library & $3.91$ ($3.99$) & $112.708$ ($6\%$) & $1.2M$ ($8.5\%$) \\
    Goodreads & $3.86$ ($3.77$) & $1.7B$ ($4.9\%$) & -- \\

\end{tabular}
 \caption{Number of readers interactions in Goodreads and Open Library. Interactions about Transnational writers are reported between parenthesis.}

 \label{table:readers}
\end{table}

Summarizing, aligning literary facts from different platforms in a unique semantic resource allows for a richer representation of World Literature, with a more balanced knowledge about Transnational writers ($+2\%$ of them are associated with at least one work). Furthermore, such data collections shows the impact of communities of readers on the diffusion of writers and their works.

\section{Visualization Platform} \label{sec:visualization}
\label{visualization_platform}
\begin{figure}[ht]
    \centering
    \includegraphics[width=\textwidth]{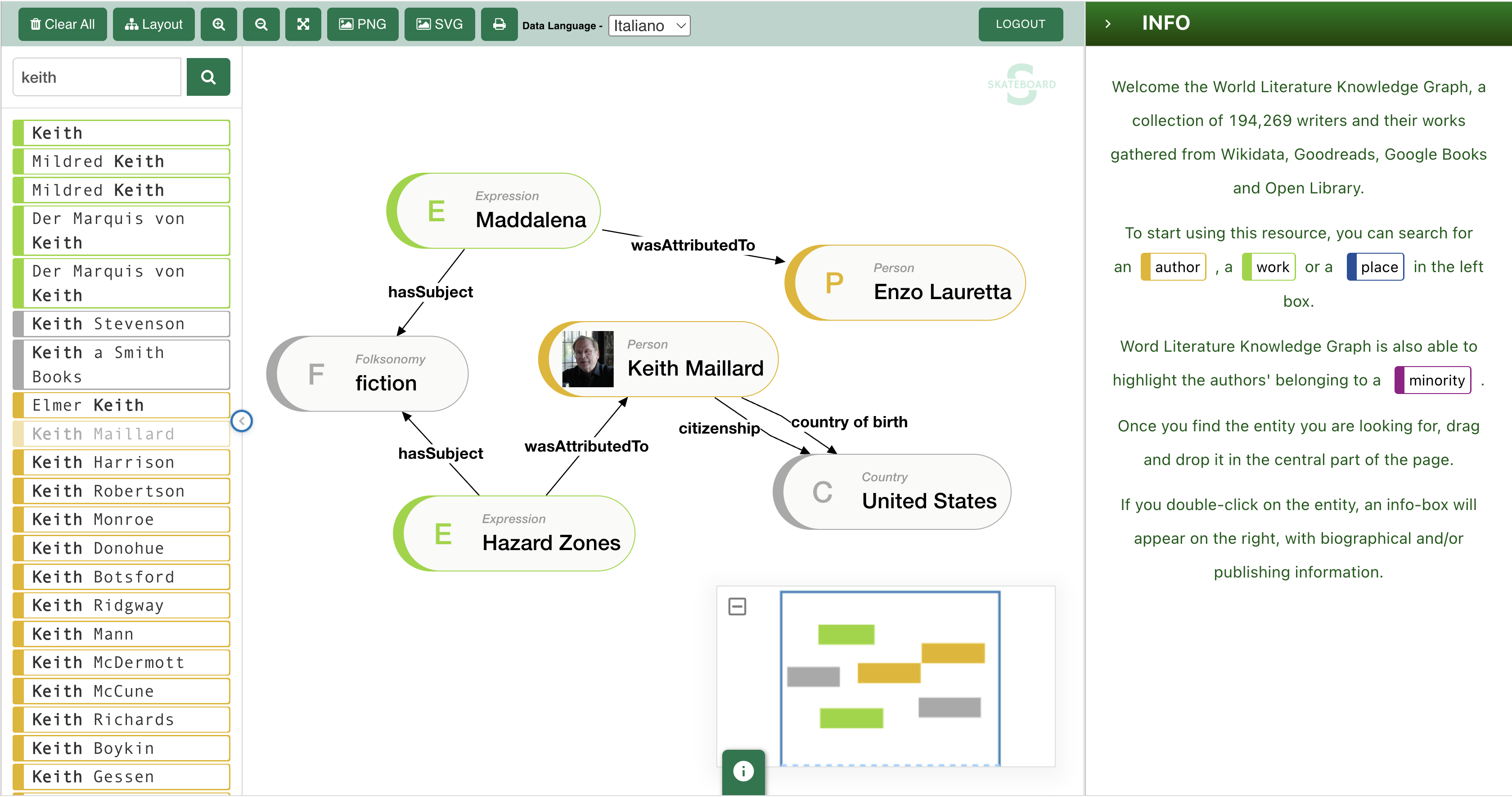}
    \caption{A snapshot of the visualization platform. On the left, the search box; in the middle, the whiteboard where entities can be dragged; on the right, info pane about the selected entity.}
    \label{fig:graph}
    \vspace{-1mm}
    
\end{figure}

\begin{sloppypar}
The World Literature Knowledge Graph is built to support advanced queries and is seamlessly integrated with SKATEBOARD, the Semantic Knowledge Advanced Tool for Extraction Browsing Organization Annotation Retrieval and Discovery, providing users with an intelligent and intuitive way to explore the vast world of literature. With the World Literature Knowledge Graph and SKATEBOARD interface, our goal is to enable users to uncover deep insights and connections within literary works and enhance their understanding of the literary world. The SKATEBOARD platform presented in this research builds upon the work of Bernasconi et al. \cite{Bernasconi2023} and represents an extension and updated version of their work to fit our specific context of use. The interface features two main views: ``Author" and ``Work". 
The navigation flow that starts with an initial search for a topic of interest.
Once a relevant topic is found, the user can drag the resource onto the central board and explore its relationships with other objects and predicates, creating a visual representation of the connections. This feature is illustrated in Figure \ref{fig:graph}.
\end{sloppypar}

\begin{figure}[ht]
    \centering
    \includegraphics[width=\textwidth]{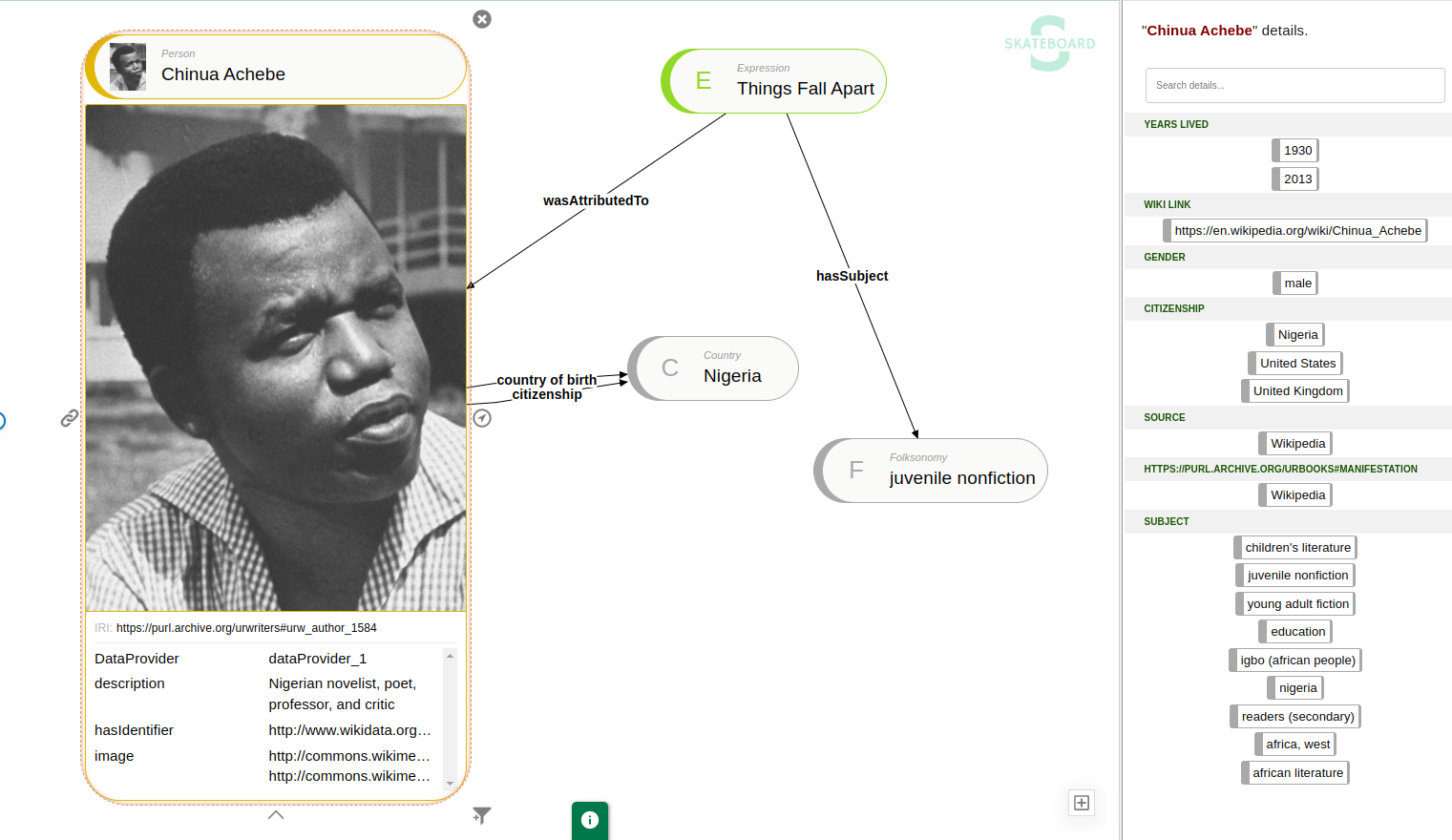}
    \caption{Person view: on the left, the central area of the interface, where  selected entities can be dragged for visualising their provenance and associated media and  their relations with other entities according to the node-link paradigm (here, Chinua Achebe); on the right, the Info pane displaying the information about the entity (e.g., biographical dates, citizenship).}
    \label{fig:author}
    \vspace{-1mm}
\end{figure}

\begin{figure}[]
    \centering
    \includegraphics[width=\textwidth]{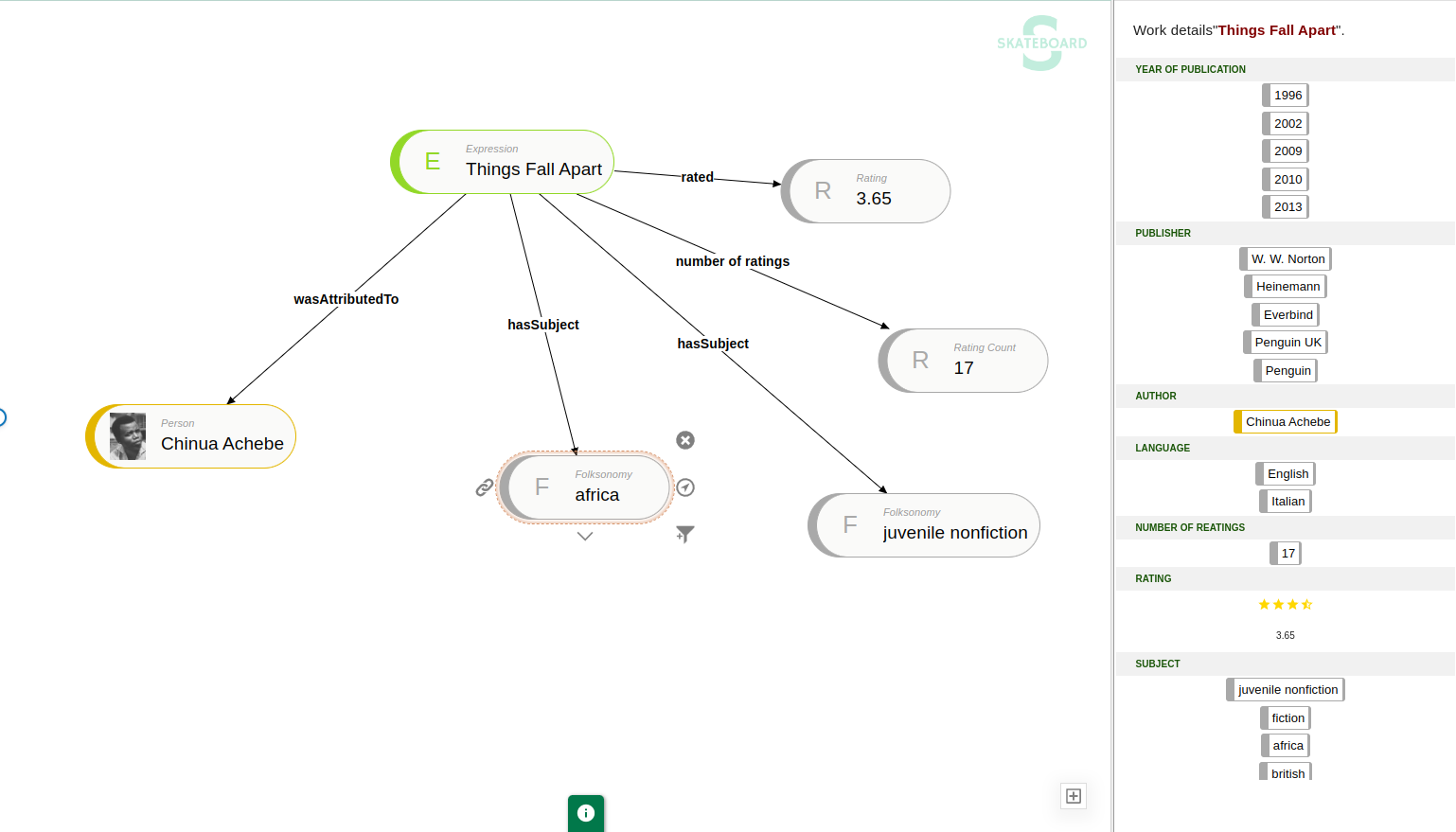}
    \caption{Expression view: on the left, the central area of the interface where a work (top left, ``Things Fall Apart") is connected with its author (Chinua Achebe, see Fig.\ref{fig:author} ). On the right, the Info pane displaying the information about the work in tabular form (an Expression in FRBR terms), such as publisher, language, rating, etc.}
    \label{fig:work}
    \vspace{-1mm}
\end{figure}

By clicking on resources of type ``Person'' (as visible in fig. \ref{fig:author}), the user can access information about an author, including both direct relationships such as published works and indirect relationships such as all the topics covered in their works, or a map of all the locations where their works were published. Clicking on resources of type ``Expression'' (as visible in fig. \ref{fig:work}) displays information specific to a particular work, such as editions, languages, and readers ratings.

Literary searches may also start from different type of entities in the Knowledge Graph. It is possible to retrieve all writers by their country of birth or by their citizenship, as well as perform searches based on specific minorities (eg.: African Americans). The platform also allows navigations based on subjects: users can browse all works linked to a specific \textsc{urb:Folksonomy}. The  graph-based  navigation encourages serendipitous discovery, allowing users to stumble upon unexpected connections and relationships.

In summary, the visualization platform presented in this research offers an updated and customizable interface for exploring and visualizing relationships between topics, authors, and works, with potential applications in various research fields.

\section{Resource Evaluation} \label{sec:evaluation}
For the evaluation of the WL-KG we organized a series of structured interviews with a group of potential targets of our resource, in line with the paradigm of user-centered design \cite{wood1997semi}: $4$ teachers, $6$ researchers in the humanities, and $3$ professionals in the publishing industry. 
Each interview was articulated in two parts: the first part, targeted on the search of Transnational writers and works, was focused on the use of the platform; the second focused on the potential uses of the resource in the users' field of work and research.

\subsubsection{Use Experience} After asking the users to search for at least one Transnational author and work of their choice, 
the use experience was investigated along three dimensions: the usability of the platform, the completeness of the results, and the accuracy of the results.

%ARCA uses two different interaction paradigms: 
%the node-link paradigm for visualising resources extracted and linked to the DBpedia knowledge base 

%the tabular paradigm for the visualisation of additional metadata related to books.

%As an information reduction strategy, ARCA allows for incremental visualisation of resources.

Concerning the usability of the platform, most users experienced  difficulties in navigating the WL-KG. First of all, they didn't realise that every element in the search area can be dragged into central whiteboard -- according to the incremental paradigm that controls the interaction between the search area and the central whiteboard (as described in Section \ref{sec:visualization}). Secondly, 
% , and subsequently explored from there, 
%relying on the platform for finding the direct connections between the entities.  
they failed to explore the information linked to the selected entity by expanding the relations between that entity and the other entities connected with it in the graph, which can be navigated in the whiteboard according the node-link paradigm (Section \ref{sec:visualization}). 
Conversely, a minority of respondents who had already experience of Knowledge Graphs found the platform easy to use and appreciated the possibility of selecting  the entities of interest by dragging them into the whiteboard, a function that they saw as a way to overcome the limitations of the standard navigation tools for graph-based representations.  
%
%\\Most users experienced some difficulties in navigating the WL-KG. 
%
Based on these observations, we hypothesize that the difficulties in the use of the visualization platform reported in the interviews can be mainly  attributed to the users' lack of experience with graph-based resources. For these users, the drag and drop selection of entities and the link-based navigation were not intuitive and can be improved by providing more guidance in the exploration (e.g., through tooltips, demo-mode navigation, etc.). 
This is in line with the comment made by some respondents who suggested to initialize the platform with an already loaded example.
Concerning the entry of the search parameters, some users expressed their difficulty in finding a suitable author or work, motivating it with their limited knowledge of the domain. To bypass this difficulty, a user suggested to create a list of writers' names, indexed by  country of birth, in a separated section of the site. We think that this suggestion is valuable, although it partly overlaps with the possibility of exploring the graph by starting from different types of entities (e.g., subjects, countries, topics), which is already available in the current version of the platform.
%Even if such issue is partially overcome by the possibility of exploring the graph from different types of entities (eg: subjects, countries), 

Concerning the completeness of the resource, a criticism derived from a misconception about its objectives shared by most respondents, who compared it with standard online archives, such as Wikipedia: the latter, being targeted at end users, include richer information about the entities in textual form, but are not suited for the development of applications that rely on the graph-based representations. 
%discovery of literary concepts. 
%
This issue can be addressed by revising the description of the resource with a clearer definition of its intended usages. A more challenging request, then, emerged from the scholars in post-colonialism,  who complained about some missing associations between works and subjects. It is the case of Andrea Levy’s work `The Long Song’: although this book is about `slavery', it is not linked to this subject in the KG, an issue  derived from the lack of attribution of this subject within the digital sources from which data were gathered.

As for completeness, almost all respondents found the resource accurate, with a few errors that we could track  from  sources. For instance, `Candide oder der Optimismus’, namely the German translation of Voltaire’s `Candide’, was attributed to Stephan Hermlin, its translator, due to an error propagated from Goodreads. To address this issues, a functionality for signaling missing and wrong information will be added in a future version of the platform. 
%in order to mitigate accuracy and completeness issues.

\subsubsection{Use Cases} The discussion of use cases was structured in two main parts: the comparison of the resource with the existing known archives and the collection of feedback about use cases and missing functionalities. Participants tended to rate the resource as useful for the discovery of new writers, but not useful for exploring new works. Such feedback reflects our data collection strategy, that was limited to the existing entities of the type writer on Wikidata and to the works that had received at least one reaction on the platforms where they are archived, aiming at relevance rather than completeness for what concerns works. 
%A similar decision led to obtain only a snapshot of these archives, which is however highly connected: users can find new writers starting from a country or from a work or a subject they master. 
Interviews also showed that almost all respondents use general purpose archives like Google, Wikipedia, and Goodreads for the literary searches,  showing a gap in the usage of knowledge bases designed for specific domains of application. The discovery of new literary facts has been pointed out as the major use case for all respondents. Interestingly, from the structured interviews with teachers, it emerged that the students themselves may be potential users of the platforms, since they could take advantage of subject-based search for supporting essay writing. Finally, it emerged the need of exposing in the knowledge base all the places where authors lived during their lives, in order to discover deeper connection between them. 

\section{Conclusion and Future Work}
In this paper we presented the WL-KG, a knowledge base of writers and works designed for the discovery of literary facts from different parts of the world and exploring the underrepresentation of non-Western writers. The resource includes $194,346$ writers and $971,120$ works collected from Wikidata, Goodreads, and Open Library. The integration of knowledge from different sources had an impact on reducing the underrepresentation of Transnational writers, about whom there is more available information in Goodreads and Open Library than in Wikidata. Our resource also allows exploring how works are received by different communities of readers.

The WL-KG is publicly available through a graph-based visualization platform that simplify its usage by non-expert users. The resource and the visualization platform were evaluated by a group of professionals whose work may be supported by the KG. Their feedback shows that the platform may be useful especially to discover new writers from multiple kinds of entities: works, subjects, countries, minority groups. Respondents also highlighted the novelty of the platform compared to existing archives: the graph-based browsing experience, designed according the link-paradigm, has been perceived as a valuable and alternative tool for exploring literary facts, even if its usability is not immediately intuitive, since graph-based resources are not widespread.      %in three fields: literary scholar, teachers, and professionals in the publishing industry. 

Future work will be devoted to improve the WL-KG with feedback emerged during the evaluation: we will increase
%implement
the knowledge base with knowledge from new communities of readers and thematic platforms; we will also release a new version of the visualization platform focused on improving its user experience for non-expert users. Finally, we will test a recommendation system based on our KG, in order to test its impact in providing fairer recommendations.

\section*{Acknowledgements}
This work was partially supported by the PNRR project CHANGES: \\Cultural Heritage Active Innovation for Next-Gen Sustainable Society, CUP H53C22000860006.

\bibliography{samplepaper}
\bibliographystyle{splncs04.bst}

\end{document}